\documentclass[12pt]{article}

\usepackage{epsf}

\begin{document}

\begin{center}

{\Large \bf Entropy concepts and DNA investigations}

\vspace{0.5cm}

Olga V. Kirillova
\vspace{0.2cm}

Department of Theoretical Physics, St.Petersburg State University

Ulyanovskaya str. 1, St.Petersburg, 198904 Russia

(e-mail:kirill@heps.phys.spbu.ru)

\end{center}

\vspace{1cm}

{\small Abstract.     
Topological and metric entropies of the DNA sequences from different
organisms were calculated. Obtained results were compared each other and
with ones of corresponding artificial sequences. 
For all envisaged DNA sequences there is a maximum of heterogeneity.
It falls in the block length interval [5,7].
 Maximum distinction between natural
and artificial sequences is shifted on 1-3 position from the
maximum of heterogeneity to the right as for metric as for topological
entropy. This point on the specificity of real DNA sequences
in the interval.}

\vspace{0.5cm}
    
PACS number(s): 87.10+e


\newpage

\section{Introduction}
One of the first conceptions of entropy was 
proposed
by C. Shannon in
application to the theory of information transmission \cite{Shan}.
Entropy in that
context implies 
the measure of heterogeneity of a set of symbols.
In mathematical form it can be written as 
$$
H=-\sum_i p_i\log p_i 
$$
where $p_i$ is probability of appearance of the $i$-th symbol. 
Then  
the notion of entropy spread 
in different fields of science:
statistical mechanics, probability theory, computer science {\it etc}.
Many
new conceptions of entropy 
such as  
metric
entropy, thermodynamic, topological, generalized, Kolmogorov-Sinai, structural spectral 
entropy
have been proposed
\cite{Badii,dentr,La, Kol, YF}.
But all of them 
pursues 
the single aim -- description of uncertainty in a
large set of objects.

One of the main directions in DNA investigations is search and elaboration of methods
for robust structural properties of genome texts extraction. What allows to understand
general principles of genetic sequences forming and makes reasonable conclusions in the
evolution theories \cite{Nu,Ei}. 
It is not 
surprising that entropy and information notions find
a wide application in DNA investigations \cite{Gatlin,Rowe,CA}. Indeed
in some sense DNA
is represented as a long sequence of symbols of the alphabet 
consisting of just four
letters. 
Eventually letter
analysis appears to be insufficient 
to explain DNA properties or structure. The
analysis of letter groups or so called words seems more interesting. But due to exponential
growth of
the number of possible words of length $n$,
when $n$ increases, it is considerably more difficult. As a rule in such
investigations some additional suggestions as, for example, about an
equidistribution of words \cite{Hp1,Hp3}, are taken. On the basis of ones
the entropy estimation is performed \cite{Hp2}. 
But in reality neither a word's distribution nor even a distribution of
letters are not equiprobable. It is essential point if we want to gain an
information from and about real DNA. Moreover very often in the analysis
short parts of a genome or chromosome have been used \cite{Hp3}. Today
available data and computer methods allow to investigate complete genomes
and chromosomes. (For example the length of human 22 chromosome is 33476902
bp.) What allows to take the maximum likelihood estimation
$p_i=q_i/N$ (where $q_i$ is the number of occurrences of the $i-$th 
word in an investigated sample) as enough good 
approximation for the calculation of the entropy in Shannon's sense. At least
in such respect an investigated object is not replaced by any artificial
set.

So we use the metric and topological definitions of entropy for DNA texts
heterogeneity estimation. 
Some remarks about the possibility of a representation of a DNA sequence
by a Markov chain for the calculation of the entropy estimate are also given.

\section{Metric entropy}
In Shannon's definition of entropy given above
$p_i$ may denote as the probability of a symbol as
the probability of a group or a block of symbols. In other words
Shannon's formula
can be rewritten as 
$$H_n=-\sum_{i=1}^{a^n} p(C_i)\log p(C_i), $$ 
where
$C_i$
is a block of symbols or 'word' of length $n$, $a$ -- the number of
letters
in a language 
and  $a^n$ -- the number of all possible combinations of length $n$ of the
letters in the language.
Obviously $H_n$ is 
non-decreasing function of $n$.
The ratio $H_n/n$ tends to a certain limit  when n goes to infinity \cite{Badii, Gatlin}
and this limit is called the metric entropy \cite{Badii}
$$
H_{met}=\lim_{n \to \infty}\frac{H_n}n .
$$

We have studied  the dependence of $h(n)\equiv H_n/n$ 
for $n\in [1,12]$ for different DNA sequences.

We used the data of the
 bacteria and yeast {\it Saccharomyces cerevisiae} complete genomes
from  GeneBank release No. 114.0 \cite{GB}
and data of 22 human complete chromosome from Sanger Centre \cite{Sanger}.

As the first criterium of the texts heterogeneity let us consider C plus G 
content (C+G) of the DNA sequences. 
It has
been obtained that
all yeast chromosomes have \% C+G
close to $38.42 \pm 0.85$. 
In the case of 20 complete bacteria
genomes C+G content varies from 29\% {\it rickettsia prowazekii (rpxx)} genome
to 65.61\% {\it mycobacterium tuberculosis (mtub)}.

It have been obtained that
for different DNA sequences reduction rate of $h(n)$ ($\Delta h(n)=h(n)-h(n+1)$) differs,
moreover
$ \Delta h(n)^{hum}<\Delta h(n)^{bact}<\Delta h(n)^{yeast}$.
The values of $h(n)$ for yeast different chromosomes are more close each other than
those for the
bacteria genomes. The chromosome IV of yeast
 and {\it mtub} bacteria genome have
minimal
difference between $h(1)$ and $h(12)$. 
First yeast
chromosome and {\it mycoplasma pneumoniae (mpneu)} bacteria
genome have maximal difference between $h(1)$ and $h(12)$.
Let us notice that the chromosome I is the shortest with maximum \% C+G 
(its length is 230204 bp). 
The chromosome IV has minimal \% C+G and it is the longest 
(its length is 1531930 bp).
The value of $h(1)$ is maximal for the
chromosome I and minimal for the chromosome IV and vice versa in case of
$h(12)$.
In Fig. 1 one can see $h(n)$ for I, IV, VI and XV  yeast chromosomes.

The bacteria genomes with greater content C+G,  contrary
to the case of  yeast, 
have minimal difference between the 
levels 1 and 12. The bacteria {\it rpxx} genome with the smallest
value of \% C+G (29\%) has the smallest $h(n)$ up to
$n=5$.
{\it Treponema pallidum (tpal)} genome has maximal value of $h(n)$ 
for $n\in [2,7]$ but it
does not hold  for
$n>7$.
In Fig. 2 the graphs of $h(n)$ for different bacteria genomes:
 ({\it rpxx}, {\it ecoli (escherichia
coli K-12 MG1655)}, {\it
tpal}, {\it mtub}, {\it mpneu}, {\it mgen (mycoplasma genitalium G37)})
are shown.

It was obtained also that genomes/chromosomes 
having approximately equal values of 
$h(1)$ and \% C+G
become fairly far from each other when the block's length increases
(see Fig. 3). What points on the genomes have distinct heterogeneity
in the 'word' context for different word's length and this can not
be found from only the letter analysis.  
It is possible to pick out three such sets from the envisaged sequences.
The first one consists of
the human 22 chromosome (\% C+G 47.8;  $h(1)$ 1.9986)
and the genomes of bacteria: {\it synechocystis PCC6803 (synecho)} (\%
C+G 47.72; $h(1)$ 1.9985),
{\it archaeoglobus fulgidus (aful)} (\% C+G 48.58; $h(1) 1.9994$).
 {\it Haemophilus influenzae Rd (hinf)} bacteria
genome (\% C+G 38.15; $h(1)$ 1.959)
and XV (\% C+G 38.16; $h(1)$ 1.9591) and XIII (\%C+G 38.2; $h(1)$ 1.9594)
yeast chromosomes form the second; the third set is presented by
{\it helicobacter pylori 26695 (hpyl)} bacteria genome (\% C+G 38.87;
$h(1)$ 1.964), VI (\% C+G 38.73;
$h(1)$ 1.963) and IX (\% C+G 38.9; $h(1)$ 1.9642) yeast chromosomes.
 The longer a sequence the slower $h(n)$ decreases.
In some aspect this observation can has a simple explanation
-- the more length of a sequence the more chances to meet
there different subsequences, in other words here the finite sample effect presents. But
one can see that for $n<6(7)$
the values of $h(n)$ for some shorter sequences are 
greater than for longer ones. This fact can not be
explained just statistically. 

\section{Topological entropy }
The topological entropy is defined as 
$$
k(n)=\frac{\log_2N(n)}{n},
$$
where
$N(n)$ is the number of distinct blocks of length $n$ in a sequence
under attention
\cite{Badii}.
We calculated $k(n)$ for $n\leq 12$.

It has been obtained that for $n\leq 4$ $k(n)$ is maximal and equal
$2$ for all genomes/chromosomes, 
i.e. all possible $n$-letters' combinations are present in even from the investigated
DNA texts.
On the fifth level just the shortest
bacteria {\it mgen} genome (its length is 580074 bp) has no some words. On
the 6 level a
half of
bacteria genomes and 7 yeast chromosomes have all possible words. (Let us notice they
are not the longer.) On the
7 level all bacteria and yeast DNA sequences have no some words. From the
8 level no all words are present even in human 22 chromosome.
Though the number of all possible words $4^8=65536$ is essentially
less of any investigated sample. Further the
$n$ increases the $k(n)$ decreases for all sequences. For yeast chromosomes
reduction takes place monotonously according to the length, the shorter a
chromosome the faster reduction.
As for the
bacteria genomes,
although here reduction also almost follows to genome length change,
it is not precisely (Fig. 4).
Thus one can say that
$k(n)$ strongly correlates with length of a sequence.

Due to check the finite size effect we generated artificial
DNA sequences of
the same size and with the same nucleotides fractions for human 22
chromosome, {\it mgen}
- the shortest bacteria genome, {\it ecoli} -- the longest (4639221 bp) bacteria genome,
{\it mtub} and {\it tpal}  bacteria genomes as well as for the longest
(IV) and the shortest (I) yeast chromosomes and compare the results
obtained from the
artificial and natural sequences.

Due to clarify the picture we have considered the differences of
$k(n)$ and $h(n)$ for artificial and natural sequences:
$\Delta^{top}_n=k(n)^{art}-k(n)^{nat}$,
$\Delta^{met}_n=h(n)^{art}-h(n)^{nat}$. Besides we have paid attention
to the difference between $k(n)^{nat}$ and $h(n)^{nat}$
($\Delta_n=k(n)^{nat}-h(n)^{nat}$). As it is easy to
see the latter value reflects heterogeneity of elements' distribution. If
only some fraction of all possible elements is realized in an envisaged
sequence and all of them are equally probable hence $k(n)=h(n)$.

As one can see the differences at beginning increase
then achieve maximum value (in the interval [7,11] for the first and
second cases and [5,9] for the third) and drop (Fig. 5, Fig. 6, Fig. 7). 
Characteristically
that maximums of the first and second cases almost coincide, the third
is shifted to the left on 1-3 positions. For the latter from envisaged
$n$ values all differences are very small and close each other (except the
human chromosome). 
This can be related just with the finite sample effect. 

\section{A DNA sequence and Markov processes}
One of the first attempts of modelling DNA sequences and obtaining
appropriate statistical characteristics based on the application of Markov
processes \cite{Gatlin, Elton}. Under the term Markov process or
Markov chain it is usually assumed the stochastic process where a state of
a system depends on its previous state. It is so called one step Markov
chain (a process with memory equal 1). In general case the memory
(dependence) may be greater than 1 but necessarily finite. In the critical
review \cite{Li97} W. Li showed that a one step Markov chain does not
characterize
observed correlation functions of DNA sequences.
We checked how considerable is supposition of Markovity for $H_n$ entropy
estimation. It is essential point because for $n>12$ natural
calculation (requiring $4^n$ values of $p(C_i)$) of $H_n$ becomes fairly
difficult. At the same time
if a process one can consider as a one step Markov chain, $p(C_i)$ can be
represented as
$$p(C)=p_{i_1}p_{i_1i_2}p_{i_2i_3}p_{i_3i_4}...p_{i_{n-1}i_n},$$
and for $H_n$ we have $$H_n=\sum_{i_1}\sum_{i_2}...\sum_{i_n}
p_{i_1}p_{i_1i_2}...p_{i_{n-1}i_n}\log[p_{i_1}p_{i_1i_2}...p_{i_{n-1}i_n}].$$
Thus for calculation of $H_n$ for any $n$ we have to know only the
transition probability matrix $\{p_{ij}\}_{i,j=1}^{a}$ and the letter
distribution $\{p_i\}_{i=1}^a$ (in our case $a=4$).
As it has been revealed the greater $n$ the greater divergence of the $H_n$ estimates 
obtained by the means of natural
calculation and ones for the Markov process.
The differences of this values for bacteria {\it ecoli}, {\it mgen}, {\it
mtub}, {\it tpal} genomes, human 22 and
yeast chromosomes IV and I for different $n$ are presented
in Table 1.

So approximation DNA by the one step Markov process for $H_n$ entropy
estimation when $n>12$ seems not to be
quite correct. At first because of progressive increase of
distinctions.

Let us note, for small
$n$ for the bacteria genomes the differences are equal. For the sequences
from distinct species the results differ even for small $n$.
As one can see here the finite sample effect is also present.

\section{Conclusion}
The effect of finiteness of a sequence seems to be considerable for large $n$ in entropy
estimation.
In average as $h(n)$ as $k(n)$ reductions correlate with length of a
sequence. 
For $n$ of order 11, 12 the differences of $h(n)$ and $k(n)$
very
small (for yeast chromosomes and the bacteria genomes) as well as the
distinctions of the natural and artificial sequences. For the human
chromosome
these distinctions are greater. 

Reduction rate of $h(n)$ for different
organisms differs. In general it is not follow to the length. Although
for yeast chromosomes for $n>8$ exactly, the longer a sequence the
greater $h(n)$. For all envisaged DNA sequences maximum of heterogeneity
has
been revealed in the interval of length $n=$[5,7]. Maximum of distinction
between natural
and artificial sequences is shifted on 1-3 position from the
maximum of heterogeneity to the right as for metric as for topological
entropy. This also points on the specificity of natural DNA sequences in the
interval $n=$[7,11].

One step Markov chain is not a good approximation for the metric 
entropy estimation. 
It is not the means to obtain a satisfactory estimate of $H_n$ when $n>12$.

Presented work confirms the fact of existence of the characteristic lengths set
revealed by spectral methods \cite{YF}. This can be related with the DNA texts
segmentation. The work is also in accordance with the block organization principle of
nucleic asids, that reflects the block-hierarchical mechanisms of evolution \cite{B}.

Possibly the investigation of a long genome (consisting of many
chromosomes) as a single sequence allows to 
achieve greater accuracy in entropy estimation and reduce the finite sample effect.

\newpage
\centerline {FIGURES}

Fig. 1 In this figure one can see $H_n/n$ versus $n$ for I, IV, VI and
XV yeast chromosomes.

Fig. 2 In this figure one can see $H_n/n$ versus $n$ for {\it rpxx},
{\it ecoli}, {\it tpal}, {\it mtub}, {\it mpneu}, {\it mgen}
bacteria genomes.

Fig. 3 In this figure one can see $H_n/n$ versus $n$ for
human 22 chromosome, genomes of bacteria: {\it synecho}, {\it aful},
{\it hinf}, {\it hpyl} and XV, XIII, VI, IX yeast chromosomes. 

Fig. 4 In this figure one can see $K_n$ versus $n$ for
human 22 chromosome, genomes of bacteria: {\it ecoli}, {\it mgen},
{\it mtub}, {\it tpal} and IV, I yeast chromosomes.

Fig. 5 In this figure one can see $\Delta_n$ (denoted D),
$\Delta_n^{met}$ (denoted Dm)
and $\Delta_n^{top}$ (denoted Dt) versus $n$ for bacteria
{\it ecoli}, {\it mgen}, {\it mtub}
genomes.

Fig. 6 In this figure one can see $\Delta_n$ (denoted D),
$\Delta_n^{met}$ (denoted Dm)
and $\Delta_n^{top}$ (denoted Dt) versus $n$ for yeast 
chromosomes I and IV.

Fig. 7 In this figure one can see $\Delta_n$ (denoted D), 
$\Delta_n^{met}$ (denoted Dm)
and $\Delta_n^{top}$ (denoted Dt) versus $n$ for human 22 chromosome.

\newpage
%

\begin{table}
\caption{In this table one can see the differences of $h(n)$
estimates obtained by the means of natural
calculation and approximation of the sequences by the one step
Markov chains for
bacteria {\it ecoli}, {\it mgen}, {\it mtub}, {\it tpal} genomes,
human 22 and yeast I and IV chromosomes for different $n$.}
\[
\begin{tabular}{cccccccc}
\hline
\multicolumn{1}{c}{n} & \multicolumn{1}{c}{mgen} &
\multicolumn{1}{c}{ecoli} & \multicolumn{1}{c}{mtub} &
\multicolumn{1}{c}{tpal} & \multicolumn{1}{c}{I} & 
\multicolumn{1}{c}{IV} & hum \\
\hline
\multicolumn{1}{c}{3} & \multicolumn{1}{c}{0.3\%} &
\multicolumn{1}{c}{0.3\%} & \multicolumn{1}{c}{0.3\%} &
\multicolumn{1}{c}{0.3\%} & \multicolumn{1}{c}{0.1\%} & 
\multicolumn{1}{c}{0.1\%} & 0.2\% \\
\multicolumn{1}{c}{4} & \multicolumn{1}{c}{0.6\%} &
\multicolumn{1}{c}{0.6\%} & \multicolumn{1}{c}{0.6\%} &
\multicolumn{1}{c}{0.6\%} & \multicolumn{1}{c}{0.25\%} & 
\multicolumn{1}{c}{0.15\%} & 0.47\% \\
\multicolumn{1}{c}{5} & \multicolumn{1}{c}{0.9\%} &
\multicolumn{1}{c}{0.9\%} & \multicolumn{1}{c}{0.9\%} &
\multicolumn{1}{c}{0.8\%} & \multicolumn{1}{c}{0.46\%} & 
\multicolumn{1}{c}{0.25\%} & 0.74\% \\
\multicolumn{1}{c}{10} & \multicolumn{1}{c}{16.96\%} &
\multicolumn{1}{c}{8.38\%} & \multicolumn{1}{c}{7.46\%} &
\multicolumn{1}{c}{16.45\%} & \multicolumn{1}{c}{26\%} & 
\multicolumn{1}{c}{13.14\%} & 5.06\% \\
\hline
\end{tabular}
\]
\label{Tab1}
\end{table}


\begin{thebibliography}{99}
%
\bibitem{Shan} Shannon C. E. Bell Syst. Tech. J. 27 (1948) 379.

\bibitem{Badii}
R. Badii, A. Politi Complexity -- hierarchical structures and scaling in physics
(Cambridge Univ. Press, Cambridge,1997).

\bibitem{dentr} Sinai Ya. G. Topics in ergodic theory (Princeton Univ.
Press, Princeton N.J. 1994).

\bibitem{La} 
L. D. Landau E. M. Lifshits Statistical physics 5 (Nauka, Moscow
1964) (in russian).

\bibitem{Kol}
A. N. Kolmogorov Problems of information transmittion
1 (1965) 3 (in russian).
\bibitem{YF} V.V. Lobzin, V.R. Chechetkin Physics Uspekhi 170 (2000) 57 (in russian).
\bibitem{Nu} R. Nussinov J. Theor. Biol. 125 (1987) 218.
\bibitem{Ei} M. Eigen Physics Uspekhi 109 (1973) 545 (in russian).
\bibitem{Gatlin}
L. Gatlin J. Theor. Biol. 10 (1966) 281.
\bibitem{Rowe}
G. W. Rowe J. Theor. Biol.101 (1983) 151.
\bibitem{CA} 
A.A.Alexandrov, N.N.Alexandrov, M.Yu. Borodovsky
et. al. Computer analysis of gene texts (Nauka, Moscow,
1990) (in russian);

Gusev V. D., Kulichkov V. A., Titkova T. N.
Calculation systems 83 (1980) 11 (in russian);

Lipman D.J., Maizel J. Nucl. Acids Res. 10 (1982) 2723.

\bibitem{Hp1} H.Herzel, A.O. Schmitt, W. Ebeling Chaos, Solitons \&
Fractals 4 (1994) 97

\bibitem{Hp3} A.O. Schmitt, H.Herzel, W. Ebeling Europhys. Lett. 23 (1993)
303

\bibitem{Hp2} H.Herzel, W. Ebeling, A.O. Schmitt Phys. Rev. E 50 (1994)
5061;

 A.O. Schmitt, H.Herzel J. Theor. Biol. 1888 (1997) 369;

I. Grosse Dynamik -- Evolution -- Strukturen (Koster Verlag, Berlin, 1996)

\bibitem{GB}URL: http://www.ncbi.nlm.nih.gov

\bibitem{Sanger}URL:
http://ftp.sanger.ac.uk/pub/human/sequences/Chr\_22/cpmlete\_sequence

\bibitem{Elton} 
R. A. Elton J. Theor. Biol. 45 (1974) 533
\bibitem{Li97} 
W. Li Computers \& Chem. 21 (1997) 257

\bibitem{B} G.R. Ivanitski, N.G. Esipova, R.A. Abagjan, S.E. Shnol Biofizika 30 (1985) 418
(in russian). 
\end{thebibliography}
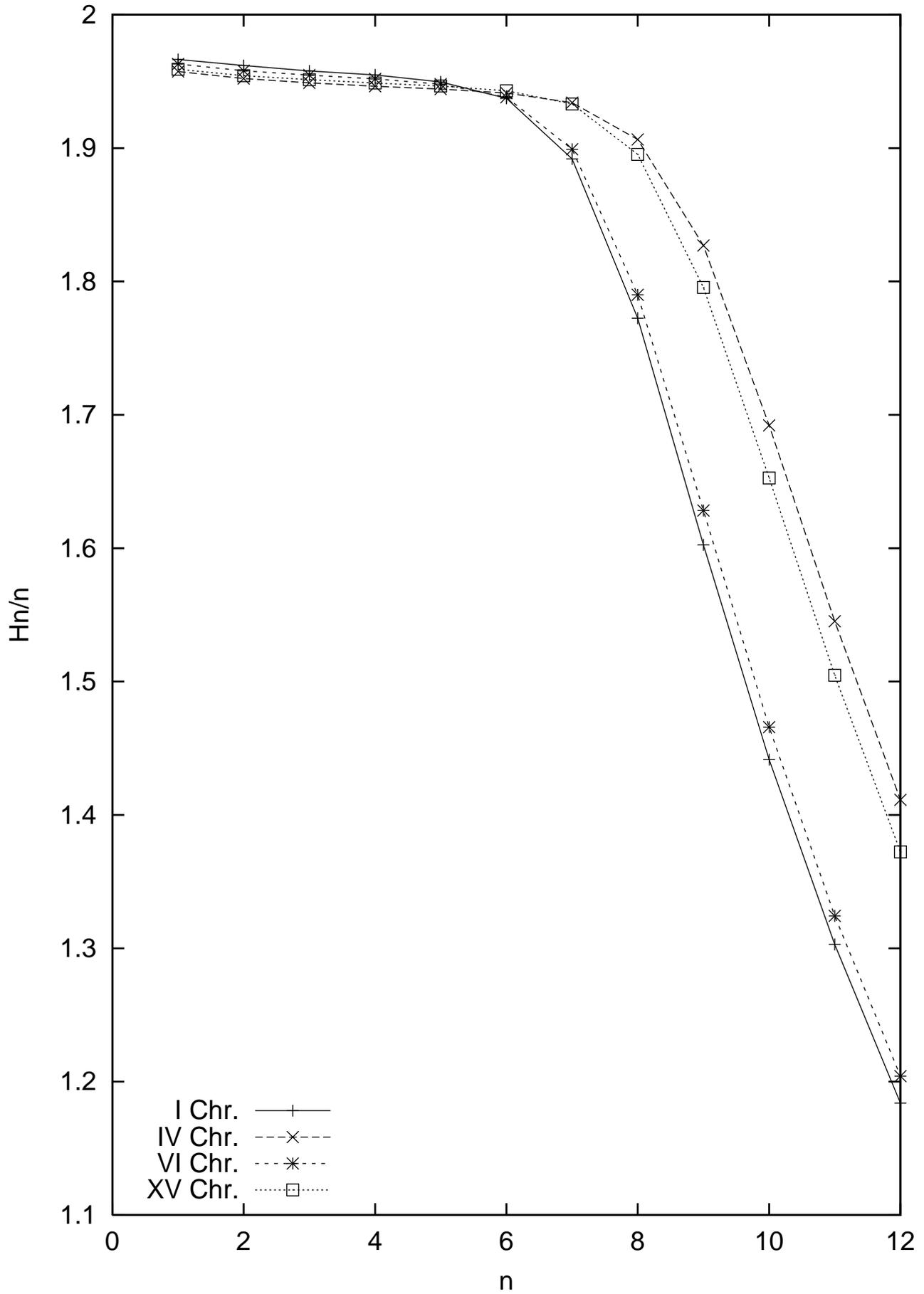
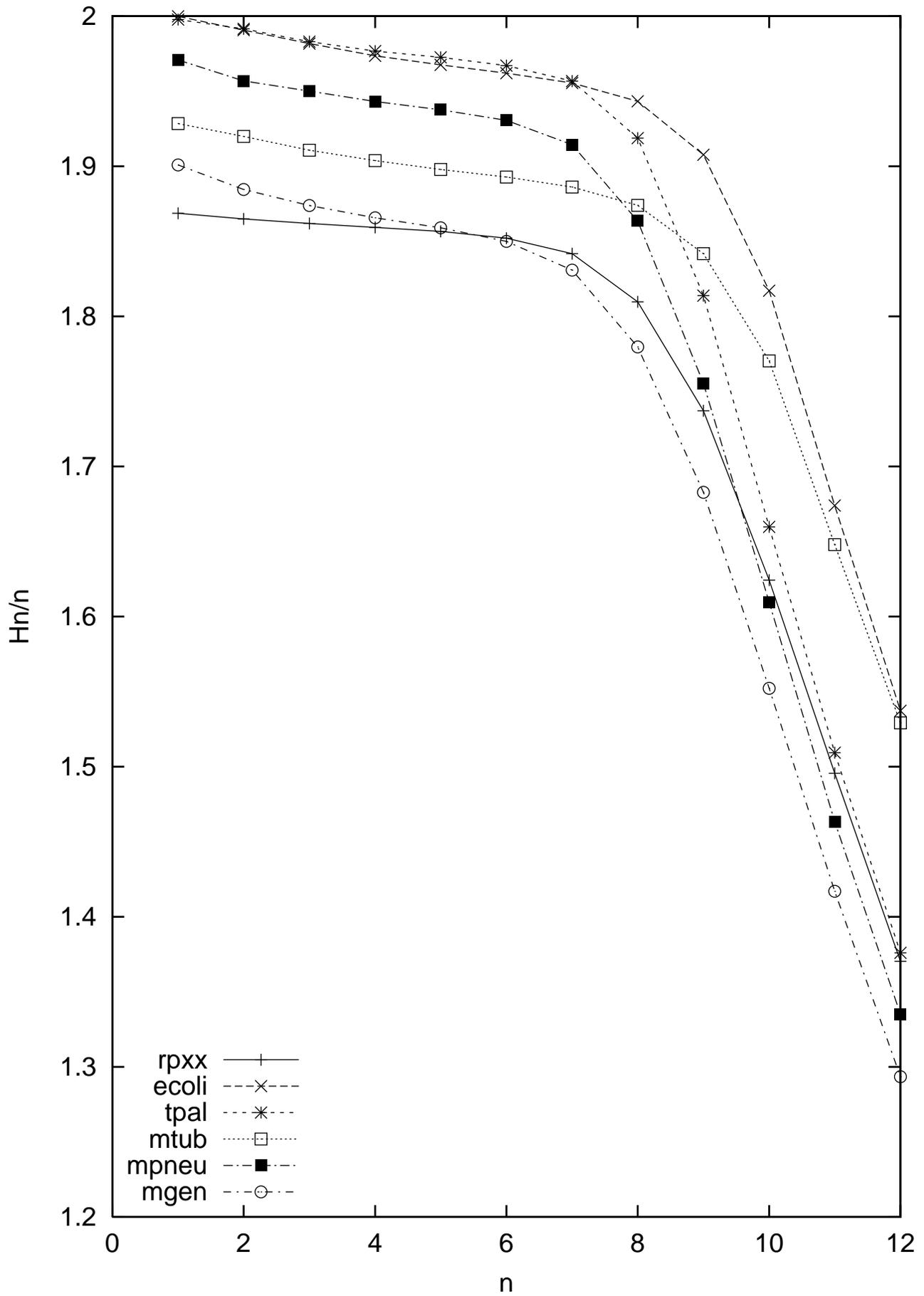
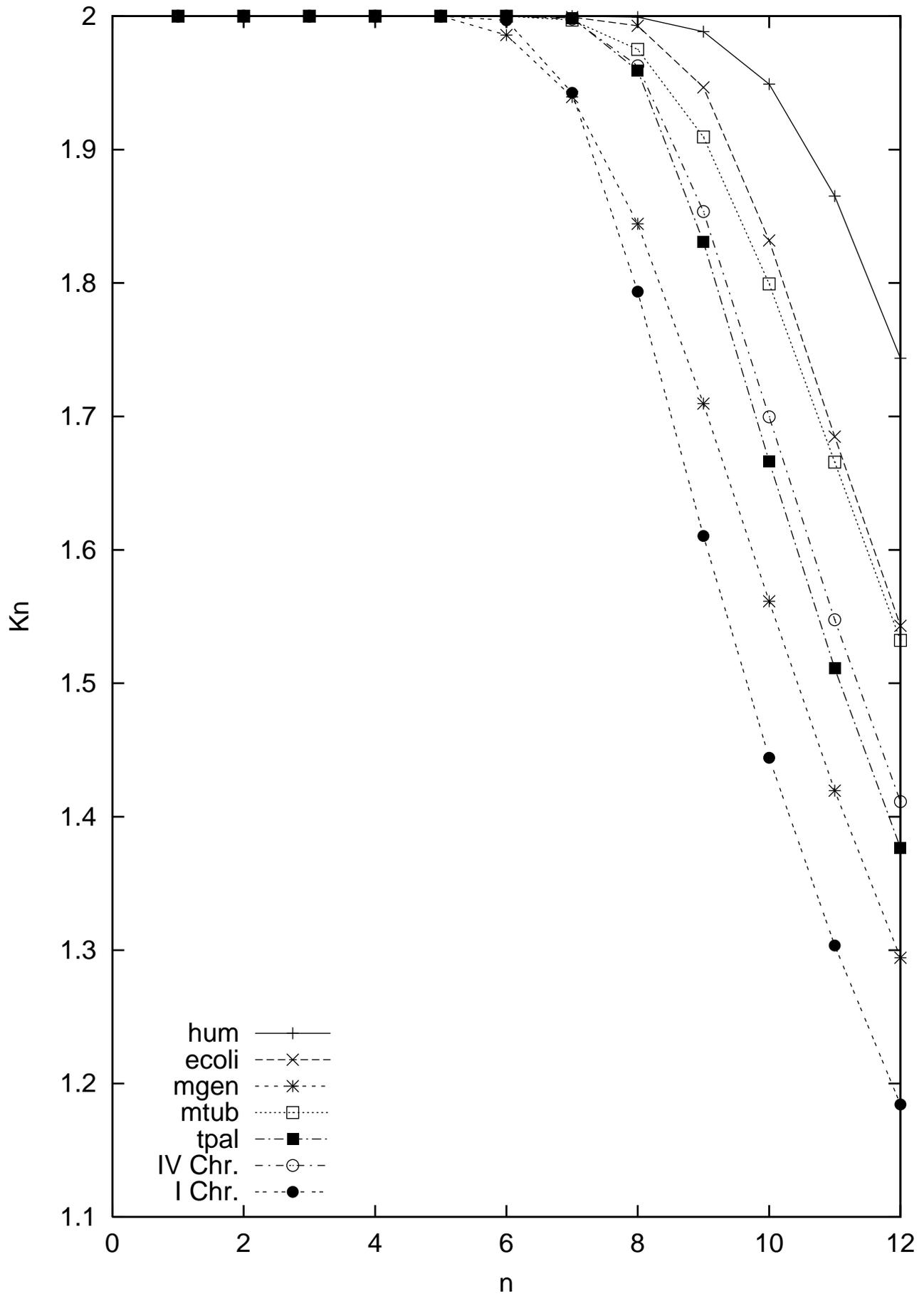
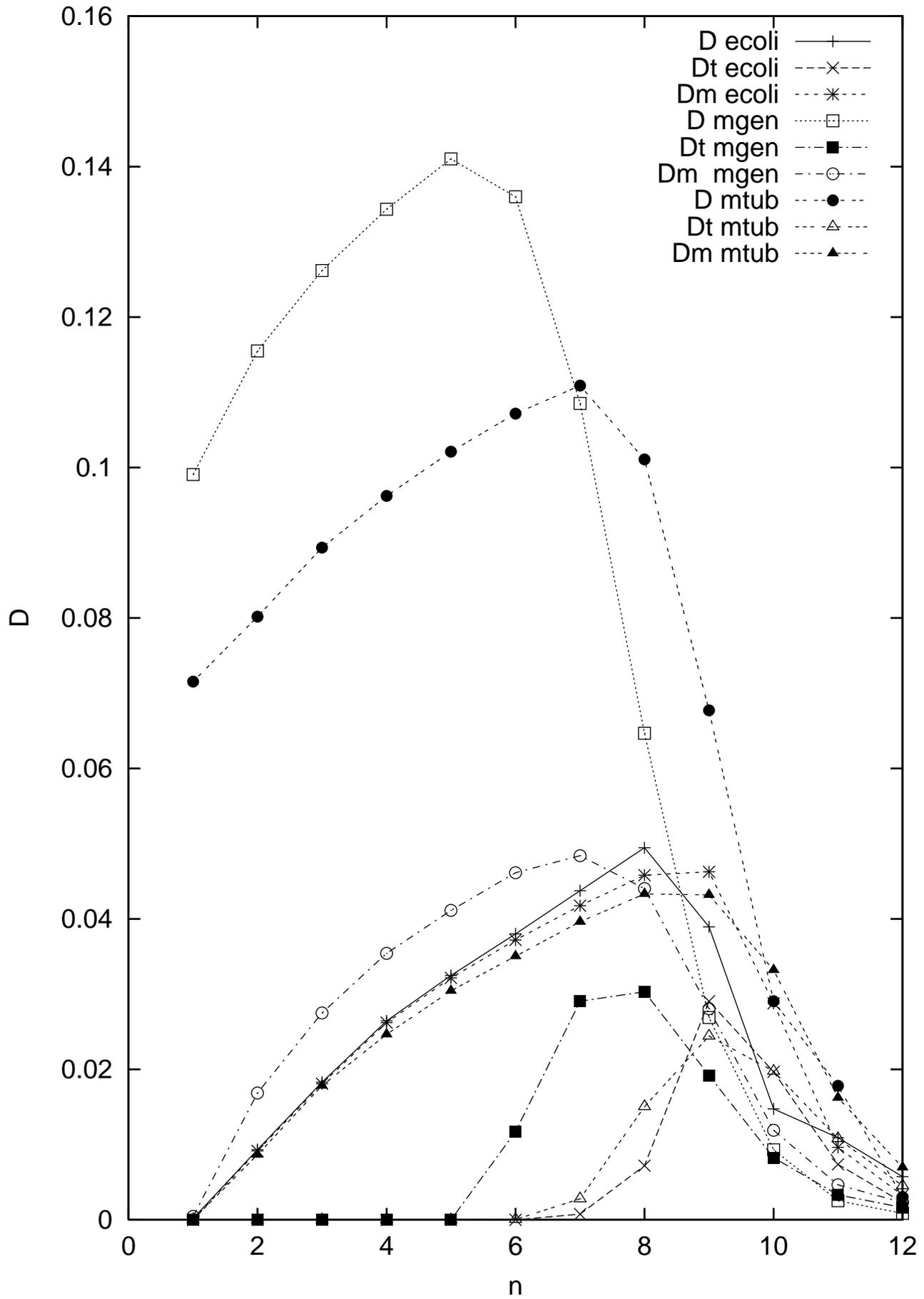
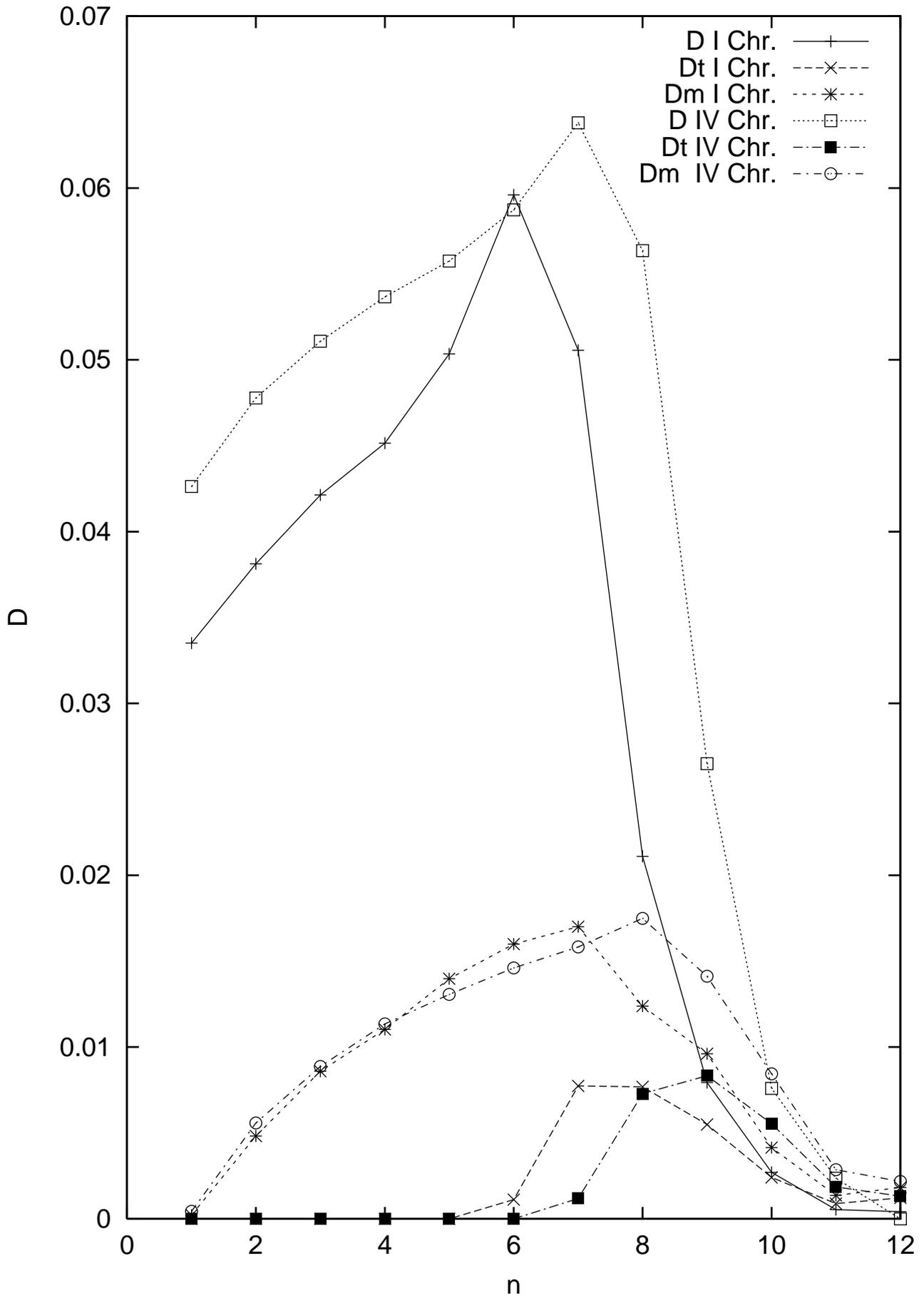
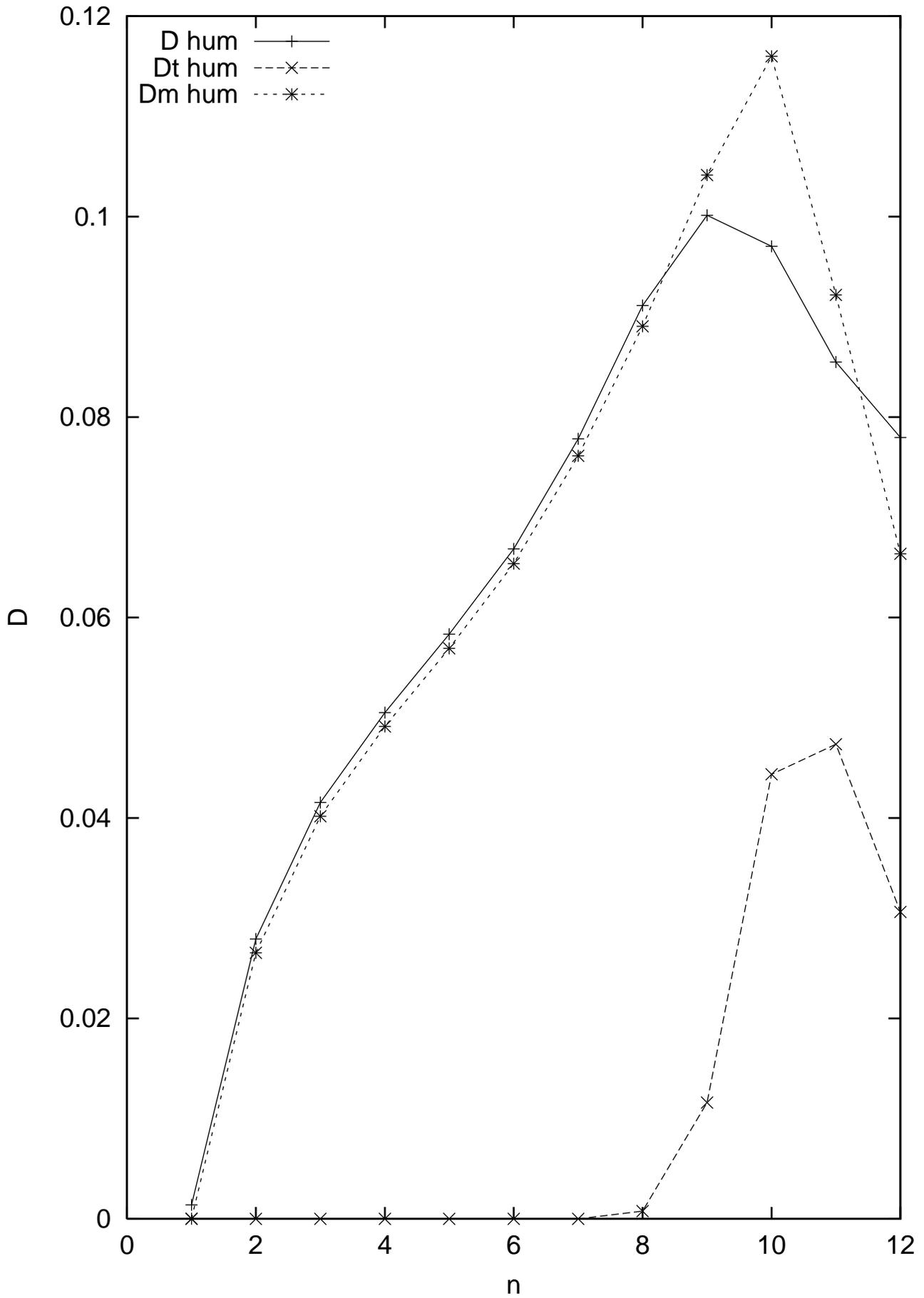
\end{document}